\documentclass[default,ms]{AGUTeX}

\usepackage{amsmath}
\usepackage{stfloats}

  \usepackage[dvips]{graphicx}
  \setkeys{Gin}{draft=false}
\authorrunninghead{HAQQ-MISRA}

\titlerunninghead{DAMPING OF GLACIAL CYCLES}

\authoraddr{Corresponding author: J. Haqq-Misra,
Blue Marble Space Institute of Science, 1200 Westlake Ave N Suite 1006, Seattle, Washington 98109, USA.
(jacob@bmsis.org)}

\begin{document}

%
%

\title{Damping of glacial-interglacial cycles from anthropogenic forcing}
%
%

%
%



\authors{Jacob Haqq-Misra,\altaffilmark{1}}

\altaffiltext{1}{Blue Marble Space Institute of Science, Seattle, Washington, USA. (jacob@bmsis.org)}





%
%


\begin{abstract}
Climate variability over the past million years shows a strong glacial-interglacial 
cycle of $\sim$100,000 years as a combined result of Milankovitch orbital forcing and 
climatic resonance. It has been suggested that anthropogenic contributions to 
radiative forcing may extend the length of the present interglacial, but the effects 
of anthropogenic forcing on the periodicity of glacial-interglacial cycles has received 
little attention. Here I demonstrate that moderate anthropogenic forcing can act to damp 
this 100,000 year cycle and reduce climate variability from orbital forcing. Future 
changes in solar insolation alone will continue to drive a 100,000 year climate cycle 
over the next million years, but the presence of anthropogenic warming can force the 
climate into an ice-free state that only weakly responds to orbital forcing. Sufficiently 
strong anthropogenic forcing that eliminates the glacial-interglacial cycle may serve 
as an indication of an epoch transition from the Pleistocene to the Anthropocene.
\end{abstract}

%
%

%

\begin{article}

%
%

\section{Introduction}

Long-term patterns of global temperature derived from ice cores \citep{petitetal,jouzeletal} 
and deep-sea oxygen isotopes \citep{shackleton} indicate cycles in the extent of global ice coverage with periods 
of 100,000, 41,000, and 23,000 years as the dominant signals \citep{haysetal,imbrieetal}. The 41,000 and 23,000 year
signals correlate directly with changes in solar insolation from variations in 
obliquity and orbital precession; however, the corresponding changes in insolation 
from eccentricity variation on a 100,000 year timescale are too weak to drive the 
glacial-interglacial features observed in the geologic record \citep{wattshayder,genthonetal,imbrieetal,rutherforddhondt,loulergueetal}. 

A possible explanation for this discrepancy between eccentricity forcing strength 
and glacial cycles is that internal climate mechanisms amplify changes in insolation 
to yield a strong $\sim$100,000 year glacial-interglacial cycle. Climatic resonances may 
occur from a combination of mechanisms, such as the thermal inertia of oceans and 
large ice sheets \citep{wattshayder,imbrieetal} or long-term cycles in 
carbon dioxide and methane \citep{genthonetal,loulergueetal}, but these 
complex feedback patterns are difficult to simulate with climate models over million 
year timescales. 

It has been suggested that anthropogenic forcing may delay the onset of the next glacial 
period \citep{mitchell,loutreberger}, but the long-term effect on the periodicity of this 
cycle has received little attention. Here a stochastic energy balance model (EBM) is used to demonstrate that anthropogenic 
forcing can damp this glacial-interglacial cycle. This EBM is tuned to a state 
of {\it stochastic resonance} that experiences variations in global average temperature with a 100,000 
year period. This periodic signal is damped with sufficiently strong 
anthropogenic forcing, which suggests the possibility that glacial-interglacial cycles 
could cease in the future.

\section{Small ice cap instability}

Simple climate models have been used to argue that ice caps can only grow to $\sim$30$^{\circ}$ latitude, 
beyond which the planet falls into a completely ice-covered state \citep{northetal,caldeirakasting}. These models also exhibit 
a {\it small ice cap instability} that result in the complete loss of an ice cap when it shrinks 
to a critical size \citep{north}. The small ice instability is of relevance to the 100,000 year 
glacial-interglacial cycle because it allows for multiple equilibrium 
solutions for present-day values of relative solar flux, 
which suggests that long-term orbital forcing could drive climate to cycle between glacial and 
ice-free conditions. 

For this study, a one-dimensional energy balance model \citep{fairenetal} is modified to calculate 
meridionally averaged temperature profiles $T$ as a function of latitude $\theta$ and time $t$ 
according to the equation
\begin{multline}
C\frac{\partial T(\theta,t)}{\partial t}=\bar{S}(\theta,L)\left(1-\alpha(T)\right)-\left(A+BT(\theta,t)\right)\\
+\frac{1}{\cos\theta}\frac{\partial}{\partial\theta}\left(D\cos\theta\frac{\partial T(\theta,t)}{\partial\theta}\right),\label{eq:EBM}
\end{multline}
where $C$ is effective heat capacity of the surface and atmosphere, $\bar{S}$ is diurnally-averaged solar 
flux, $\alpha$ is top-of-atmosphere albedo, $A$ and $B$ are infrared flux constants, and $D$ is a diffusive 
parameter to describe the efficiency of meridional energy transport. Diurnally-averaged 
solar flux $\bar{S}=Sf(\theta,L)$ is the product of a constant solar flux $S$ and a function of latitude $\theta$ and orbital 
longitude $L(t,e,{\delta}_0,L_p)$, which depends on present-day values of eccentricity $e$, obliquity ${\delta}_0$, 
and the longitude of perihelion $L_p$, to yield seasonally-varying insolation \citep{gaidoswilliams}. Uniform geography 
is assumed with constant heat capacity $C = 2.1\times10^8$ J m$^{-2}$ K$^{-1}$ to simulate an aquaplanet.
Albedo is defined as $\alpha = 0.291$ for $T \ge 263.15$ and $\alpha = 0.663$ for $T < 263.15$ (note that the precision of 
these albedo values reflects the tuning of the model \citep{caldeirakasting,gaidoswilliams}.) The parameter $D$ is set to a fixed value $D = 0.38$ W m$^{-2}$ K$^{-1}$ 
that allows the model to reproduce present-day Earth conditions \citep{gaidoswilliams}. Outgoing infrared flux is 
parameterized as a linear function of temperature with constants $A = 203.3$ W m$^{-2}$ and $B = 2.09$ 
W m$^{-2}$ $^{\circ}$C$^{-1}$ tuned to present-day Earth \citep{northetal}. The model is discretized into 18 equally spaced latitudinal zones with an 
initial temperature profile $T(\theta,0) = T_0(\theta)$ and stepped 
through $\tau$ complete orbits by increments of ${\Delta}t = 8.64\times10^3$ s to 
numerically solve Eq. (\ref{eq:EBM}).

\begin{figure}[h]
\noindent\includegraphics[width=20pc]{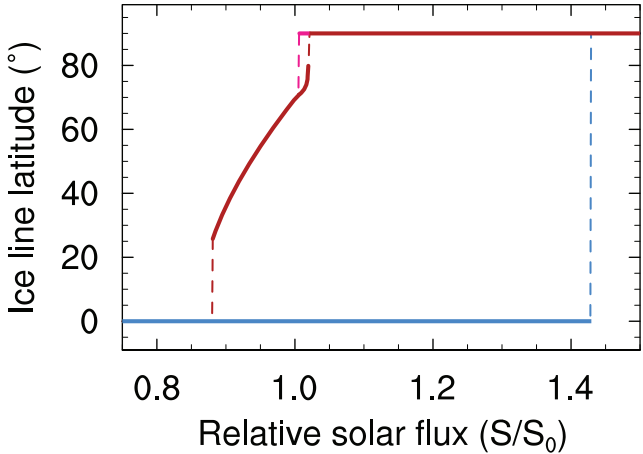}
\caption{Energy balance model for the climate of Earth. Equilibrium climate states are calculated 
by initializing a deterministic energy balance model from present-day (red), ice-free (pink), and 
ice-covered (blue) conditions. Dotted lines show discontinuous transitions between climate states, 
including a maximum ice line of ${\sim}25^{\circ}$ and a deglaciation threshold of $S/S_0 > 1.4$. 
A “small ice cap instability” is evident as discrete transitions between ice-free and ice-cap 
states (dashed red and pink lines near $S/S_0 = 1$), which permits multiple equilibrium solutions 
for certain values of $S/S_0$.}
\label{fig:bistability}
\end{figure}

This model behavior is shown in Fig. \ref{fig:bistability},  which gives the equilibrium ice line latitude as a 
function of relative solar flux. These equilibrium climate states are calculated by 
initializing the EBM with present-day Earth ($T_0(|\theta|>70^{\circ}) < 263.16$ K and 
$T_0(|\theta|<70^{\circ}) > 263.16$ K, red curve), ice-free ($T_0(\theta) > 273.16$ K, pink curve), and ice-covered 
($T_0(\theta) = 233$ K, blue curve) conditions for each value of relative solar flux $S/S_0$ (where $S_0 = 
1360$ W m$^{-2}$). The EBM is iterated for $\tau = 100$ model years in increments of ${\Delta}\tau = 1$ yr to reach 
a statistically steady-state (where ${\delta}T/{\delta}t{\approx}0$) and find the extent of the ice line latitude ${\theta}_i$ 
(where $T({\theta}_i) = 263.16$ K). This results in 
two hysteresis loops: one between ice-covered and ice-free states (from ${\sim}0.85 < S/S_0 < {\sim}1.45$) 
and another between small ice cap and ice-free states (near $S/S_0 {\sim}1.0$). This second loop is 
the small ice cap instability that describes transitions between glacial and ice-free conditions.

This simplified model is designed to highlight the multiple equilibria that can exist for Earth-like 
climates. Ice cap instabilities are well-known features of EBMs \citep{northetal,north}, 
and more recent studies with global climate models (GCMs) have shown some success in demonstrating a maximum 
ice line latitude and a large ice cap instability \citep{decontopollard,ishiwatarietal}. The small ice cap instability 
has also been observed in GCM simulations \citep{leenorth,winton}, although it is not yet clear 
why this feature is present in some GCMs but not others. With this in mind, the use of an EBM in the analysis that follows is 
intended to be a qualitative approach to a problem that eventually could be approached with 
more sophisticated tools.

\section{Stochastic resonance}

The concept of {\it stochastic resonance} suggests that small random fluctuations in temperature 
are amplified by non-linear responses in the climate system, which establishes periodic 
transitions between glacial and ice-free conditions when orbital forcing is applied \citep{benzietal,imkeller}. 
These random fluctuations represent the cumulative effect of any climatic resonance 
mechanisms present in the Earth system and allow for an abstracted implementation of 
resonance in climate models. A climate model in a state of stochastic resonance, when 
driven by Milankovitch orbital forcing, will show a 100,000 year periodicity in the 
global extent of glacial ice (ranging from large ice caps to ice-free conditions) associated 
with a change in global average temperature (spanning a range of ${\sim}10$ K), in addition to the 
shorter glacial cycles. 

\begin{figure}[b]
\noindent\includegraphics[width=20pc]{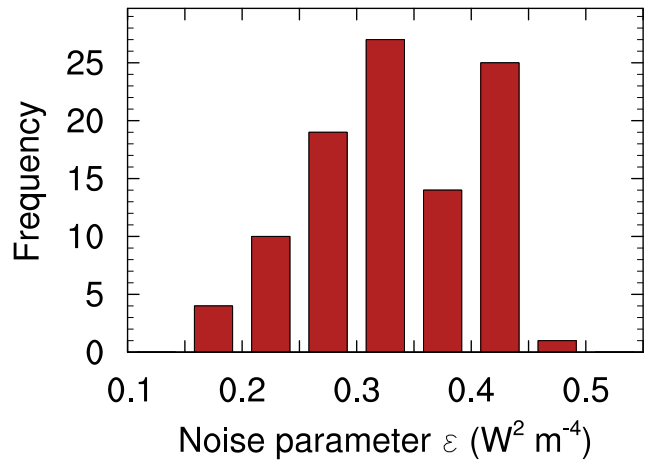}
\caption{Histogram of estimated values for the noise parameter $\varepsilon$. This distribution 
includes 100 values with mean 0.334 W$^2$ m$^{-4}$ and standard deviation 0.07 W$^2$ m$^{-4}$.}
\label{fig:histogram}
\end{figure}

A stochastic EBM is developed by modifying Eq. (\ref{eq:EBM}) to include a randomly fluctuating term, so that 
\begin{multline}
C\frac{\partial T(\theta,t)}{\partial t}=\bar{S}(\theta,L)\left(1-\alpha(T)\right)-\left(A+BT(\theta,t)\right)\\
+\frac{1}{\cos\theta}\frac{\partial}{\partial\theta}\left(D\cos\theta\frac{\partial T(\theta,t)}{\partial\theta}\right)+\sqrt{\varepsilon}\dot{W},\label{eq:stochEBM}
\end{multline}
where $\varepsilon$ is a tunable constant known as the "noise parameter", and $\dot{W}$ represents Gaussian white noise 
with zero expectation and unit variance. Eq. (\ref{eq:stochEBM}) represents a stochastic EBM when 
$\varepsilon > 0$ and reduces to a deterministic EBM (Eq. (\ref{eq:EBM})) when $\varepsilon = 0$.

The constant $\varepsilon$ in Eq. (\ref{eq:stochEBM}) is optimized to find a state of stochastic 
resonance \citep{benzietal,imkeller} so that the EBM exhibits periodic transitions between warm and cool climates 
while avoiding global glaciation. Variations in mean annual insolation between -1 Myr and +1 
Myr from the present are obtained from orbital and precessional calculations \citep{laskaretal} that yield 
time-dependent quantities for solar flux $S(\tau)$, eccentricity $e(\tau)$, obliquity $\delta_0(\tau)$, and the 
longitude of perihelion $L_p(\tau)$. This allows the stochastic EBM to iterate for $\tau = 2$ million 
model years in increments of ${\Delta}\tau = 1000$ years to simulate periodic variability of past and 
future climate. An optimal choice of $\varepsilon$ will exhibit glacial cycles near the 100,000, 41,000, 
and 23,000 yr periods that result from orbital and precessional forcing \citep{imbrieetal}, whereas this 
feature of stochastic resonance will be lost if $\varepsilon$ is too small or too large \citep{benzietal,imkeller}. 

An appropriate choice for $\varepsilon$ is a value that maximizes a power spectrum (or periodogram) 
at a 100,000 yr period. A golden section search algorithm is applied to the stochastic EBM 
to identify a value of $\varepsilon$ that yields a power spectrum 
with a maximum value closest to 100,000 yr. This optimization algorithm is initialized with a search 
interval for $\varepsilon$ between 0.1 and 0.5 W$^2$ m$^{-4}$, and two sets of 2 Myr stochastic 
EBM calculations are performed with two different values of $\varepsilon$ within this range. 
The power spectrum of each of these model calculations is evaluated at a period of 100,000 yr, 
and the calculation that gives a larger value of this 100,000 yr power spectrum signal is retained for further optimization. 
Iteration continues until an optimal value of $\varepsilon$ is reached with a tolerance 
of 0.001 W$^2$ m$^{-4}$. Following a sample average approximation method \citep{kleywegtetal}, 
this optimization process is repeated 100 times, and the resulting set of 100 estimated 
values is shown as a histogram in Fig. \ref{fig:histogram}. The average of 
this distribution of gives an expected optimal value of $\varepsilon = 0.334$ W$^2$ m$^{-4}$ with 
a standard deviation of 0.07 W$^2$ m$^{-4}$.

\begin{figure*}[b]
\noindent\centering\includegraphics[width=30pc]{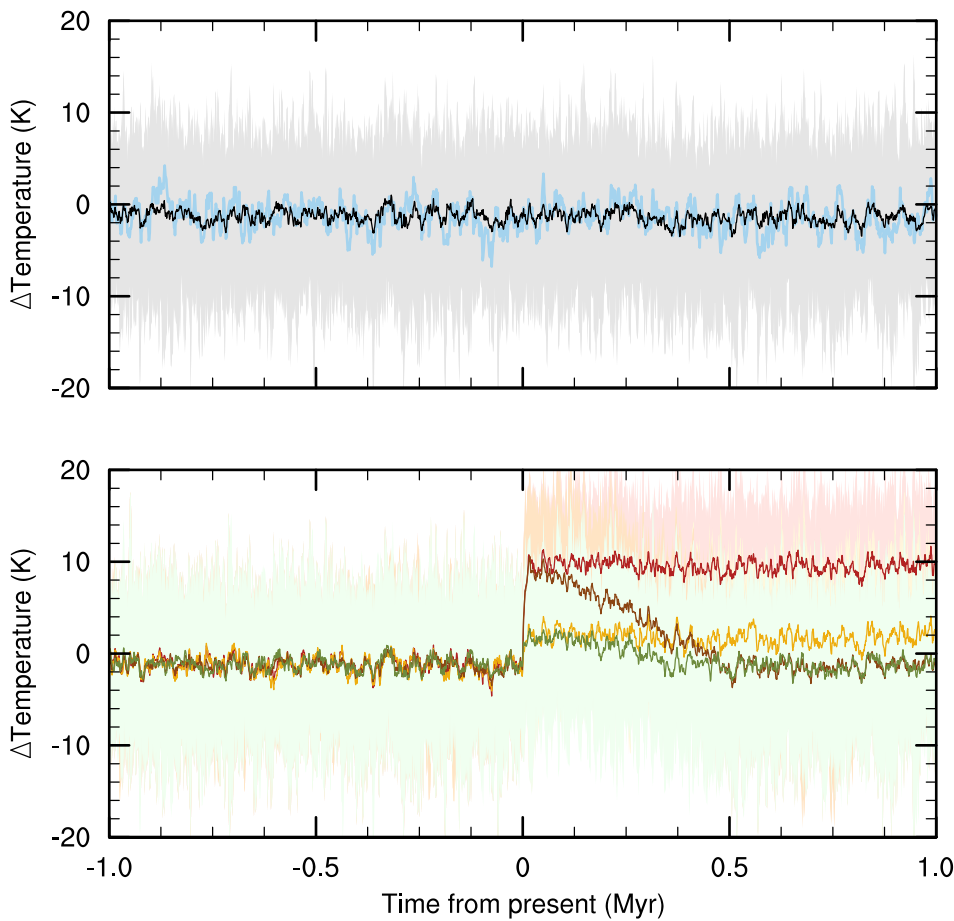}
\caption{Temporal evolution of global average temperature from -1 Myr to +1 Myr.
The change in global average temperature from the present is calculated with a 
stochastic energy balance model ensemble as a function of past and future solar 
insolation. The control ensemble (top panel) includes no anthropogenic forcing and 
maintains a state of stochastic resonance between glacial and inter-glacial 
conditions. The black curve shows the full 100-member ensemble mean, the blue curve shows 
a 10-member ensemble mean, and the gray shaded area indicates the range of 
ensemble members. Anthropogenic ensembles (bottom panel) include scenarios with 
moderate (gold) and extreme (red) changes to radiative forcing at 0 Myr that continue 
indefinitely. The gradual recycling of greenhouse gases is represented by additional 
scenarios with moderate (green) and exterme (brown) forcing that 
include a linear reduction in greenhouse gas forcing from 0 to +0.5 Myr. Shaded areas indicate 
the range of ensemble members.}
\label{fig:temps}
\end{figure*}

This tuned stochastic EBM is driven by Milankovitch orbital forcing \citep{laskaretal} 
to calculate global average temperature from -1 Myr to +1 Myr in the absence of any anthropogenic 
forcing. A 100-member control ensemble of 2 Myr stochastic 
EBM calculations (with $\varepsilon$ set to the expected optimal value) is calculated to give an 
expectation of global temperature evolution, and Fig. \ref{fig:temps} (top panel) shows the full 100-member ensemble 
mean (black curve), a 10-member ensemble mean (blue curve), and the range of ensemble members (gray shading).
The system is in stochastic resonance between climates warmer and cooler than today and manages to 
avoid global ice-covered or permanent ice-free states that would halt this periodicity. 

\begin{figure*}[t]
\noindent\includegraphics[width=40pc]{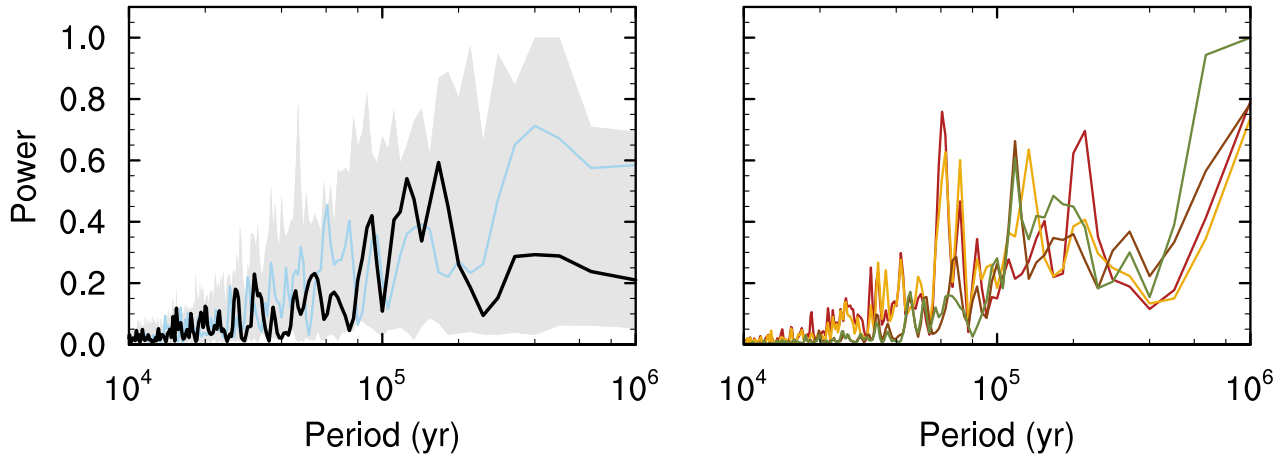}
\caption{Normalized power spectrum of long-term climate variability. 
The control ensemble (left panel) includes no anthropogenic forcing, 
and the full 100-member ensemble mean shows its strongest peaks at $\sim$100,000 to $\sim$180,000 yr (black curve). 
The blue curve shows the power spectrum for a 10-member ensemble mean, and the gray shaded area indicates 
the range of ensemble members. Anthropogenic ensembles (right panel) include moderate (gold, green) and extreme 
(red, brown) forcing under scenarios of indefinite emissions (gold, red) and gradual reduction (green, brown). In 
all of these anthropogenic scenarios, the $\sim$100,000 yr periodicity is damped in comparison to more dominant signals with 
larger periods.}
\label{fig:power}
\end{figure*}

An expected power spectrum of this 100-member control ensemble is shown in Fig. \ref{fig:power} (left panel). 
The full 100-member ensemble (black curve) shows its strongest peaks at $\sim$100,000 to $\sim$180,000 yr,
which coincides approximately with the strongest signal of climate variability observed 
during the past 1 Myr. The 10-member ensemble subset (blue curve) has a more pronounced signal at 
400,000 yr, although a peak near a $\sim$100,000 yr period is still evident, and the range of ensemble 
members (gray shading) indicates peaks at both $\sim$200,000 and $\sim$400,000 yr. (Note that all curves in 
the left panel of Fig. \ref{fig:power} are normalized by the maximum ensemble power.)
Although deterministic replication of these cycles is beyond the scope of this study, these stochastic EBM calculations
suggest that long-term periodicity should continue into the future if additional forcing is absent.

\section{Anthropogenic forcing}

Anthropogenic forcing is represented by changing the outgoing infrared radiative flux to simulate 
the effect of increased greenhouse gases on climate. The first scenario of moderate forcing is analogous 
to increasing atmospheric carbon dioxide by a factor of two, which is represented by reducing the outgoing 
infrared radiative flux parameter $A$ by  $3.7$ W m$^{-2}$ (an estimated value of the radiative forcing due 
to doubled carbon dioxide \citep{ipcc}) when $\tau = 0$ Myr (present-day). A doubling of carbon dioxide occurs in many climate model predictions under 
a variety of emissions scenarios, with the doubling occurring in 100 to 200 years or less \citep{ipcc}, so this scenario 
of moderate forcing may be within the scope of probable outcomes. Likewise, the second scenario of extreme forcing 
corresponds to an eightfold increase in atmospheric carbon dioxide by reducing the parameter $A$ by $14.8$ W m$^{-2}$ 
(resulting from four doublings of carbon dioxide) at $0$ Myr. This extreme forcing scenario corresponds approximately to consuming 
nearly all fossil fuel reserves, which would have catastrophic consequences for humanity. The likelihood of 
such an extreme scenario may be small or unknown, but in this study it serves to indicate an upper limit on the degree of 
anthropogenic forcing.

Carbon dioxide is slowly recycled by the carbonate-silicate cycle on timescales of $\sim$0.5 Myr or longer 
\citep{walkeretal,berneretal}, which suggests two additional scenarios for consideration. After the 
moderate or extreme change in anthropogenic forcing, a gradual reduction is applied to the infrared parameter 
$A$ over a period of 0.5 Myr as a representation of recycling by the carbonate-silicate cycle. This parameterization 
admittedly is crude, as neither radiative forcing nor geologic recycling can be adequately 
represented as linear processes. Nevertheless, these scenarios are intended to bracket the range of possible 
outcomes for long-term changes to the climate system.

Similar to the control, 100-member ensembles are calculated for these four anthropogenic forcing scenarios.
Global average temperature for these scenarios is shown in Fig. \ref{fig:temps} (bottom panel) and includes 
moderate (gold curve, green curve) and extreme (red curve, brown curve) anthropogenic forcing under scenarios of 
indefinitely continuing emissions (gold curve, red curve) and gradual reduction of emissions by the 
carbonate-silicate cycle (green curve, brown curve). The dark curves show the full 100-member ensemble mean, and 
light shading indicates the range of ensemble members. These scenarios exhibit a sharp upward trend at 0 Myr 
when anthropogenic forcing begins, and the resulting evolution from 0 Myr to 1 Myr shows a reduction in the 
amplitude of temperature fluctuations. Although oscillating temperatures continue into the future, 
the amplitude of these changes appears to be at least somewhat reduced relative to the control scenario.

This damping of the $\sim$100,000 yr glacial cycle is evident from the power spectra of these 
anthropogenic forcing scenarios in Fig. \ref{fig:power} (right panel). These spectra are normalized by the 
maximum power of each scenario, and the least squares linear trend is removed from the scenarios with gradual 
reduction due to the carbonate-silicate cycle. In all scenarios, the most dominant signal is at longer periods near 
$~\sim$1 Myr, with much weaker signals in the range of $\sim$100,000 to $\sim$200,000 yr. Shorter-period 
variability associated with changes in obliquity and precession are still evident in most of these scenarios, 
but anthropogenic forcing  appears to be sufficiently strong to damp the $\sim$100,000 glacial-interglacial cycle 
even with gradual recycling by the carbonate-silicate cycle. These calculations illustrate the potential for 
anthropogenic forcing to delay or halt the glacial-interglacial cycle; however, 
identifying the exact threshold of forcing at which this will occur is beyond the scope of the stochastic 
EBM employed here.

\section{Conclusion}

These calculations suggest the existence of a threshold for anthropogenic forcing, beyond 
which the climatic response to Milankovitch orbital forcing will be damped and the 100,000 
year glacial-interglacial cycle will cease. Identifying this threshold will require the use of 
sophisticated atmospheric-oceanic general circulation models coupled to ice sheet models in order 
to approach deterministic, rather than stochastic, predictions about future changes to ice age cycles. 
Nevertheless, the simpler calculations presented here provide a robust and computationally efficient 
method for demonstrating anthropogenic effects on climate variability.

If long-term anthropogenic forcing is relatively 
weak or if climate sensitivity is low, then the onset of the next glacial cycle may be delayed 
by $\sim$50 kyr \citep{mitchell,loutreberger}. But with stronger anthropogenic forcing or high climate sensitivity, the 
cessation of glacial-interglacial cycles will indicate a permanent transition to the geologic epoch of the Anthropocene.


%

\begin{acknowledgments}
The data for this paper is available upon request from the author.
Funding for this research was provided by the NASA Astrobiology 
Institute's Virtual Planetary Laboratory (award no. NNX11AC95G,S03).
\end{acknowledgments}




%

%
%
\end{article}
%
%
%
%
%
%

%
%


\end{document}